\documentclass[sigconf,screen,balance,authorversion]{acmart}
\setcopyright{none}
\AtBeginDocument{%
  }

\usepackage{comment}
\usepackage{xspace}
\usepackage{lipsum}
\usepackage{multirow}
\usepackage{tabulary}
\usepackage{pgfplots}
\usepackage{pgfplotstable}
\usepackage{enumitem}

\newcommand{\BfPara}[1]{\noindent\textbf{#1.}\xspace}

\usetikzlibrary{pgfplots.groupplots,arrows,automata,positioning,arrows.meta,patterns,shapes.geometric,arrows.meta,spy,shapes.multipart,bending,fit}
\usepgfplotslibrary{dateplot,fillbetween,statistics,colormaps,colorbrewer}
\pgfplotsset{compat=1.18}

\newcommand{\arr}{\tikz[baseline]  \draw[thick, gray!90!black, arrows={-latex[scale=1.5, color=black, round]}] (0, 0.09) to[bend left=0] (0.40, 0.09);} 

\pgfplotsset{
  tick label style = {font=\sffamily},
  every axis label = {font=\sffamily},
  legend style = {font=\sffamily},
  label style = {font=\sffamily},
  title style = {font=\sffamily},
}
\tikzset{every picture/.style={/utils/exec={\sffamily}}}

\pgfplotsset{
    restrict x to domain**/.code args={#1:#2}{
        \pgfkeysalso{/pgfplots/x coord trafo=#1}
        \let\numericxmin\pgfmathresult
        \pgfkeysalso{/pgfplots/x coord trafo=#2}
        \let\numericxmax\pgfmathresult
        \pgfkeysalso{/pgfplots/restrict x to domain={\numericxmin}:{\numericxmax}}
    }
}

\newenvironment{customlegend}[1][]{%
    \begingroup
    \csname pgfplots@init@cleared@structures\endcsname
    \pgfplotsset{#1}%
    }{%
    \csname pgfplots@createlegend\endcsname
    \endgroup
}%

\def\addlegendimage{\csname pgfplots@addlegendimage\endcsname}

\pgfplotscolorbarCMYKworkaroundfalse

\pgfplotsset{
  nodes near coords greater equal only/.style={
    small value/.style={
      /tikz/coordinate,
    },
    every node near coord/.append style={
      check for small values/.code={
        \begingroup
        \pgfkeys{/pgf/fpu}
        \pgfmathparse{\pgfplotspointmeta<#1}
        \global\let\result=\pgfmathresult
        \endgroup
        \pgfmathfloatcreate{1}{1.0}{0}
        \let\ONE=\pgfmathresult
        \ifx\result\ONE
          \pgfkeysalso{/pgfplots/small value}
        \fi
      },
      check for small values,
    },
  },
}

\newcommand{\eg}{\textit{e.g.}}
\newcommand{\ie}{\textit{i.e.}}

\tikzstyle{nodestyle} = [very thick, fill opacity=0.8, minimum size=15pt, minimum width=15pt, minimum height=15pt,
font=\small\sffamily, text=black, inner sep=2pt, outer sep=2.5pt]

\tikzstyle{textnode} = [
    font=\footnotesize\sffamily, text=black,
    inner sep=0pt, draw=none, align=center,
    fill=white, fill opacity=0.8,
]

\tikzstyle{motiftextnode} = [
    textnode, fill=none,
]

\tikzstyle{main_victim} = [nodestyle, regular polygon, regular polygon sides=4, draw=YellowGreen, fill=YellowGreen!30, scale=1]

\tikzstyle{aggressive_victim} = [nodestyle, regular polygon, regular polygon sides=4, draw=ProcessBlue, fill=ProcessBlue!30]

\tikzstyle{non_aggressive_victim} = [nodestyle, regular polygon, regular polygon sides=4, draw=Goldenrod, fill=Goldenrod!50]

\tikzstyle{non_agg_defender_victim} = [nodestyle, regular polygon, regular polygon sides=5, draw=magenta, fill=magenta!50]

\tikzstyle{bully} = [nodestyle, circle, minimum size=10pt, draw=Orange, fill=Orange!70]

\tikzstyle{bully_asst} = [nodestyle, circle, minimum size=10pt, draw=Red, fill=Red!70]

\tikzstyle{aggressive_defender} = [nodestyle, regular polygon, regular polygon sides=3, draw=Orchid, fill=Orchid!50]

\tikzstyle{non_agg_defender_bully} = [nodestyle, regular polygon, regular polygon sides=3, draw=Salmon, fill=Salmon!50]

\tikzstyle{non_agg_defender_victim} = [nodestyle, regular polygon, regular polygon sides=3, draw=magenta, fill=magenta!50]

\tikzstyle{motifnode} = [minimum size=10pt, minimum width=10pt, minimum height=10pt,
font=\small\sffamily, text=black, inner sep=2pt, outer sep=2.5pt]

\tikzstyle{defender} = [motifnode, regular polygon, regular polygon sides=3, draw=Orchid, fill=Orchid, scale=1.10]

\tikzstyle{victim} = [motifnode, diamond, draw=YellowGreen!50!Green, fill=YellowGreen!50!Green, scale=1.35, outer sep=1.5pt]
\tikzstyle{bullyy} = [motifnode, circle, draw=Orange!80!Red, fill=Orange!80!Red, scale=0.90]  

\tikzstyle{motifedge} = [thick, tips=proper, draw=Gray!50!black, -{Triangle[color=black]}]

\tikzstyle{motifedge_light} = [tips=proper, draw=CadetBlue!80!Gray, -{Triangle[color=CadetBlue!50!Gray]}]
\tikzstyle{motifedge_heavy} = [line width=1.6pt, tips=proper, draw=BrickRed, -{Triangle[color=BrickRed!80, scale=0.70]}]

\tikzstyle{edgestyle} = [
thick, 
tips=proper, draw opacity=0.4
]

\tikzstyle{heavy} = [
    edgestyle, 
    line width=2pt,
]

\tikzstyle{main_vic2agg_def} = [
edgestyle, 
draw=YellowGreen!80!black, fill=YellowGreen!50, 
-{latex[fill=YellowGreen!80!black]}]

\tikzstyle{vic2agg_def} = [edgestyle, draw=BlueViolet, fill=BlueViolet!50, -{latex[fill=BlueViolet!80!black]}]

\tikzstyle{bully2vic} = [edgestyle, draw=RedOrange, fill=RedOrange!50, -{latex[RedOrange, fill=RedOrange!95!black]}]

\tikzstyle{bully2main_vic} = [edgestyle, draw=RedOrange, fill=RedOrange!50, -{latex[RedOrange, fill=RedOrange!80!black]}]

\tikzstyle{bully2agg_victim} = [bully2vic]

\tikzstyle{agg_vic2bully} = [edgestyle, draw=ProcessBlue!70!black, fill=Aquamarine!50, -{latex[Aquamarine, length=5pt, fill=Cyan!70!black]}]

\tikzstyle{agg_vic2agg_def} = [edgestyle, draw=Cyan, fill=Cyan!50, -{latex[Cyan, length=5pt, fill=Cyan!70!black]}]

\tikzstyle{non_agg_vic2agg_def} = [edgestyle, draw=Dandelion, fill=Dandelion!50, -{latex[Dandelion, length=5pt, fill=Dandelion!70!black]}]

\tikzstyle{agg_def2bully} = [edgestyle, draw=Plum, fill=Plum!50, -{latex[Plum, length=5pt, fill=Plum!80!black]}]

\tikzstyle{non_agg_def2bully} = [edgestyle, draw=Salmon, fill=Salmon!50, -{latex[Salmon, length=5pt, fill=Red!80!black]}]

\tikzstyle{non_agg_defA2bully} = [edgestyle, draw=magenta, fill=magenta!50, -{latex[magenta, length=5pt, fill=magenta!80!black]}]

\tikzstyle{agg_vic2bully_asst} = [edgestyle, draw=ProcessBlue, fill=ProcessBlue!50, -{latex[ProcessBlue, length=5pt, fill=ProcessBlue!80!black]}]

\tikzstyle{vic2bully} = [
edgestyle, 
draw=YellowGreen!80!black, fill=YellowGreen!50, 
-{latex[fill=YellowGreen!80!black]}]

\tikzstyle{vic2def} = [
edgestyle, 
draw=YellowGreen!80!black, fill=YellowGreen!50, 
-{latex[fill=YellowGreen!80!black]}]

\tikzstyle{def2bully} = [edgestyle, draw=Plum, fill=Plum!50, -{latex[Plum, length=5pt, fill=Plum!80!black]}]

\tikzstyle{bully2def} = [edgestyle, draw=Orange, fill=Orange!50, -{latex[Orange, length=5pt, fill=Orange!80!black]}]

\tikzstyle{inline} = [scale=0.40]
\tikzstyle{AggDef} = [aggressive_defender, inline]
\tikzstyle{NonAggDefBully} = [non_agg_defender_bully, inline]
\tikzstyle{Bully} = [bully, inline]
\tikzstyle{AggVic} = [aggressive_victim, inline]

\begin{document}

\title{Network Analysis of Cyberbullying Interactions on Instagram}

\author{Satyaki Sikdar}
\orcid{0000-0003-1669-6594}
\affiliation{
\institution{Loyola University Chicago}
\city{Chicago}
\state{IL}
\country{USA}
}
\email{ssikdar@luc.edu}

\author{Manuel Sandoval}
\orcid{0009-0009-8590-8811}
\affiliation{
\institution{Loyola University Chicago}
\city{Chicago}
\state{IL}
\country{USA}
}
\email{msandovalmadrigal@luc.edu}

\author{Taylor Hales}
\orcid{0009-0007-7827-1109}
\affiliation{
\institution{Loyola University Chicago}
\city{Chicago}
\state{IL}
\country{USA}
}
\email{thales@luc.edu}

\author{Chloe Kilroy}
\orcid{0009-0002-7149-6931}
\affiliation{
\institution{Loyola University Chicago}
\city{Chicago}
\state{IL}
\country{USA}
}
\email{ckilroy@luc.edu}

\author{Maddie Juarez}
\orcid{0009-0002-6701-4469}
\affiliation{
\institution{Loyola University Chicago}
\city{Chicago}
\state{IL}
\country{USA}
}
\email{mjuarez4@luc.edu}

\author{Tyler Rosario}
\orcid{}
\affiliation{
\institution{Loyola University Chicago}
\city{Chicago}
\state{IL}
\country{USA}
}
\email{trosario@luc.edu}

\author{Juan J. Rosendo}
\orcid{0009-0008-6284-8468}
\affiliation{
\institution{Loyola University Chicago}
\city{Chicago}
\state{IL}
\country{USA}
}
\email{jrosendo@luc.edu}

\author{Deborah L. Hall}
\orcid{0000-0003-2450-3596}
\affiliation{
\institution{Arizona State University}
\city{Glendale}
\state{AZ}
\country{USA}
}
\email{d.hall@asu.edu}

\author{Yasin N. Silva}
\orcid{0000-0003-1852-1683}
\affiliation{
\institution{Loyola University Chicago}
\city{Chicago}
\state{IL}
\country{USA}
}
\email{ysilva1@luc.edu}

\renewcommand{\shortauthors}{Sikdar et al.}

\begin{abstract}
Cyberbullying continues to grow in prevalence and its impact is felt by thousands worldwide. This study seeks a network science perspective on cyberbullying interaction patterns on the popular photo and video-sharing platform, Instagram. Using an annotated cyberbullying dataset containing over 400 Instagram posts, we outline a set of heuristics for building Session Graphs, where nodes represent users and their cyberbullying role, and edges represent their exchanged communications via comments. Over these graphs, we compute the Bully Score, a measure of the net malice introduced by bullies as they attack victims (attacks minus pushback), and the Victim Score, a measure of the net support victims receive from their defenders (support minus attacks). Utilizing small subgraph (motif) enumeration, our analysis uncovers the most common interaction patterns over all cyberbullying sessions. We also explore the prevalence of specific motif patterns across different ranges of Bully and Victim Scores. We find that a majority of cyberbullying sessions have negative Victim Scores (attacks outweighing support), while the Bully Score distribution has a slight positive skew (attacks outweighing pushback). We also observe that while bullies are the most common role in motifs, defenders are also consistently present. This suggests that bullying mitigation is a recurring structural feature of many interactions. To the best of our knowledge, this is the first study to explore this granular scale of network interactions using human annotations at the session and comment levels on Instagram.
\end{abstract}



\acmDOI{}
\acmYear{2025}
\acmConference{}{}{}

\settopmatter{printacmref=false, printccs=true, printfolios=true}


\maketitle

\begin{figure}[ht]
    \centering
\pgfdeclarelayer{background}
\pgfdeclarelayer{foreground}
\pgfsetlayers{background,main,foreground}

\tikzstyle{bully_mp} = [bully, circle split, font=\scriptsize\sffamily, inner sep=1.5pt]

\tikzstyle{main_victim_mp} = [main_victim, rectangle split, rectangle split parts=2, font=\scriptsize\sffamily, inner ysep=1.5pt]
\tikzstyle{aggressive_victim_mp} = [aggressive_victim, rectangle split, rectangle split parts=2, font=\scriptsize\sffamily, inner ysep=1.5pt]
\tikzstyle{non_agg_defender_bully_mp} = [non_agg_defender_bully, rectangle split, rectangle split parts=2, font=\scriptsize\sffamily]

\pgfplotsset{
    legend image with multipart/.style={
        legend image code/.code={%
            \node[#1, semithick, anchor=center, font=\tiny\sffamily, scale=1.2, inner sep=0.5pt, inner ysep=0.75pt] at (0.3cm,0cm) {out\nodepart{lower}in\nodepart{two}in};
        }
    },
}
{
\begin{tikzpicture}

    \begin{pgfonlayer}{foreground}
        


        \node[main_victim_mp, scale=1.25] (mv1) at (0, 0) {3.27 \nodepart{two} 4.80};
        \node[aggressive_victim_mp, scale=1.25] (av1) at (-1.75, 0) {3.20 \nodepart{two} 1.80};
        \node[aggressive_victim_mp, scale=1.25] (av2) at (1.75, 0) {4.20 \nodepart{two} 0};

        \node[bully_mp] (b1) at (-2.5, 2.50) {2 \nodepart{lower} 7.27};
        \node[bully_mp] (b2) at (0.0, 3.25) {1 \nodepart{lower} 7.27};
        \node[bully_mp] (b3) at (2.5, 2.50) {3.60 \nodepart{lower} 2.40};
        
        \node[aggressive_defender, scale=1.25] (ad1) at (-2.5, -2.50) {};
        \node[aggressive_defender, scale=1.25] (ad2) at (0, -3.25) {};
        \node[non_agg_defender_bully, scale=1.25] (nadb1) at (2.5, -2.50) {};

        \node[text=OrangeRed!80!Red, align=left, font=\small] at (-1.85, -11.5em) {Bully Score: $-$3.44};
        
        \node[text=TealBlue!80!black, align=left, font=\small] at (2, -11.5em) {Victim Score: $+$1.35};
    \end{pgfonlayer}

    \begin{pgfonlayer}{background}
        \draw[main_vic2agg_def, line width=0.89] (mv1) edge (ad1);  
        \draw[main_vic2agg_def, line width=0.91] (mv1) edge (ad2);  
        \draw[bully2vic, line width=1] (b1) edge (mv1);  
        \draw[bully2vic, line width=0.71] (b2) edge[bend right=5] (mv1);  
        
        \draw[agg_def2bully, line width=0.89] (ad1) edge[bend left=15] (b1);  
        \draw[agg_def2bully, line width=0.89] (ad1) edge[bend left=25] (b2);  
        
        \draw[agg_def2bully, line width=0.91] (ad2) edge[bend right=10] (b1);  
        \draw[agg_def2bully, line width=0.91] (ad2) edge[bend right=11] (b2);  

        \draw[agg_vic2bully, line width=0.89] (av1) edge (b1);  
        \draw[agg_vic2bully, line width=0.89] (av1) edge (b2);  
        
        \draw[bully2vic, line width=0.95] (b3) edge (av1);  
        \draw[bully2vic, line width=0.95] (b3) edge (mv1);  
        
        \draw[agg_vic2bully, line width=0.84] (av2) edge[bend right=20] (b1);  
        \draw[agg_vic2bully, line width=0.84] (av2) edge[bend right=10] (b2);  

        \draw[non_agg_def2bully, line width=0.71] (nadb1) edge[bend right=17] (b1);  
        \draw[non_agg_def2bully, line width=0.71] (nadb1) edge[bend left=3] (b2);  
        \draw[agg_vic2bully, line width=0.84] (av2) edge (b3);  
        
        \draw[non_agg_def2bully, line width=0.71] (nadb1) edge (b3);  
    \end{pgfonlayer}
    \begin{scope}[shift={(9em, -16.5em)}]
        \begin{customlegend}[
            legend columns=0,
            legend style={
            draw=black,
            fill=none,
            font=\scriptsize,
            column sep=0.4em,
            row sep=0.1em,
            cells={align=left},
          },
          legend cell align={left},
          legend entries={Bully~~,{Main\\ Victim},{Agg\\Victim},{Agg\\ Defender},{Non-Agg\\Defender}}
          ]
            
            \addlegendimage{legend image with multipart=bully_mp, mark options={scale=1}}
            \addlegendimage{legend image with multipart=main_victim_mp}
            \addlegendimage{legend image with multipart=aggressive_victim_mp}
            \addlegendimage{aggressive_defender, mark=triangle*,  mark options={scale=1.75}, only marks}
            \addlegendimage{non_agg_defender_bully, mark=triangle*,  mark options={scale=1.75}, only marks}
        \end{customlegend}
    \end{scope}
\end{tikzpicture}
}
    \caption{Session Graph constructed from comment annotations in Tab.~\ref{tab:annotated-comments}.
    Certain nodes are presented with their weighted out (top) and in (bottom) degrees. Victim nodes are denoted as rectangles, and Bully nodes as circles. The Bully and Victim scores are the difference between the weighted out- and in-degrees (top $-$ bottom) averaged over all Bully and Victim nodes, respectively.}
    \label{fig:session-digraph}
\end{figure}



\section{Introduction}
Social interactions define the human experience. 
Positive interactions help foster connections, build communities, and enrich our day-to-day lives. 
There is also a darker underbelly of communication---bullying, hate speech, name calling, and other means of verbal abuse---that are rife with harmful intent.
These behaviors present differently based on the communication medium and the user demographics.
In this work, we present insight into the negative and often harmful interactions found on online social media platforms from \textit{cyberbullying}. 
Despite lacking a concrete definition, cyberbullying is often recognized as an aggressive, intentional act carried out by an individual or a group of individuals using electronic forms of contact, repeatedly and over time against a victim who cannot easily defend themselves~\cite{the_og,what_is_bullying}.
For example, a 2022 survey conducted by the Pew Research Center found that 46\% of teens ages 13 to 17 in the US experienced at least one form of cyberbullying~\cite{pewresearchTeensCyberbullying}.

The diversity of user demographics along with the multifaceted nature of social media platforms adds complexity when attempting to understand and analyze cyberbullying ~\cite{cb_multilevel_analysis,global_trends_on_research,cb_a_bibliometric_analysis}. 
Furthermore, finding large quantities of cyberbullying data for analysis is challenging due to the limited availability of labeled datasets, the high cost of labeling at both session- and comment-level, and strict query limitations imposed by most social media platforms (allowing access to only recent posts, or only posts from users with public profiles and large numbers of followers). Given these constraints, we worked with a previously collected Instagram dataset with cyberbullying annotations at the session level and integrated granular comment-level annotations. These annotations including topics, severity, and user role.

In this study, we utilize the aforementioned annotations to provide a network science focused analysis to patterns of cyberbullying on Instagram.  
Our objectives can be summarized as follows:
\begin{itemize}
    \item Construct localized user interaction networks (graphs) using labeled comments from Instagram posts, hence referred to as sessions, with prior evidence of cyberbullying. 
    The proposed network construct is a novel representation of Instagram user interactions based on their cyberbullying roles. 
    Fig.~\ref{fig:session-digraph} provides a small example Session Graph from the dataset.
    
    \item Quantify and characterize the different behavioral tendencies of the users across Instagram sessions based on metrics derived from the aforementioned networks. 
    Specifically, we define the \textit{Bully Score} and \textit{Victim Score}. 
    Conceptually, these scores track the residual flow of malice and peer support accumulated on the Bully and Victim nodes, respectively.
    
    \item Investigate the different \textit{motifs} (also known as graphlets)---connected sub-graphs with three or four nodes---that arise in the Session Graphs. 
    We tabulate the frequently occurring motifs to reveal interesting mesoscale interaction patterns between groups of actors. 
\end{itemize}

\section*{Related Work}
\BfPara{Cyberbullying \& Social Science}
Cyberbullying is a complex interpersonal behavior with features and characteristics that are distinct from its offline counterpart~\cite{cb_regular_b_overlap,meta_cb_overlap}. 
For instance, with traditional bullying, the bullying interactions are confined to a physical location. 
With cyberbullying, interactions escape their physical bounds and can manifest  in whichever way the bully can contact or make a statement about the victim using electronic means.
Cyberbullying is also highly context dependent, manifesting differently based on the specific characteristics of the participants involved, developmental stages, and location~\cite{changes_in_latent,longitudinal_patterns_cb}. 
While cyberbullying is often characterized as occurring primarily between two participants, it is shaped by individual, family, peer, community, and cultural  factors~\cite{understanding_the_psychology,related_factors}. 
Prior research has found cyberbullying typically involves different behavioral roles from its participants.
Five commonly explored roles in the literature are: \textit{Bully}, \textit{Victim}, \textit{Bully Supporter}, \textit{Defender of the Victim}, and \textit{Bystander}.
We provide a detailed description of various cyberbullying roles relevant to this work in Tab.~\ref{tab:bullying-roles}.

Incorporating user roles alongside severity is key to better grasp the persistent and pervasive nature of cyberbullying~\cite{perp_victim_cb,cb_past_present_future}.
An important nuance is that user roles are seldom \textit{static} within a session. This is especially true if session participants make multiple comments within a session. 
Behavioral actions also permeate across sessions wherein past role incarnations leave downstream effects. Specially, the act of cyberbullying is a known predictor of prior victimization and vice-versa~\cite{POVEDANO201544,duality_of_cb}.

\BfPara{Cyberbullying \& Network Analysis}
Networks, also referred to as graphs, are a flexible construct to model complex, interconnected systems where entities have relationships. 
Networks are widely used by sociologists to model complicated phenomena, \eg, political polarization on social media and homophily in social networks~\cite{conover2011political,christakis2013social}. 
Relevant to our discussion is the application of networks in the context of cyberbullying. 
Users within a session are a natural choice for nodes in a network with edges documenting their interactions. 
The specific characteristics of the edges, such as their directionality and weight, depend on specific context and nuances of the interactions.
For example, directed edges can be used to model one-sided, directional relationships.
Edge weights, on the other hand, can be used to signify the frequency of interactions, or encode the intensity associated with these interactions~\cite{squicciarini_identification_2015,soleimanian_network_2024}. 
Networks have also been used to draw connections between an individual's aggressive behavior in relation to someone's social status~\cite{faris_status_2011}.
In prior work, researchers have also turned to networks as input to cyberbullying classifiers. 
For example, \citet{wang_cyberbullying_2021} constructed follower networks that captured the strength of relationships between community members. 
The work by \citet{kao_understanding_2019} is particularly relevant for its consideration of cyberbullying roles (Victim, Bully, and Supporter).
In their work, \citet{kao_understanding_2019} identified victim-bully and victim-supporter pairs as found in their user-comment ego networks. 
These pairs were then utilized for classifying individual user comments. 

\section{Data}

In this section, we introduce our data source, a collection of curated Instagram sessions with known occurrences of cyberbullying.
We will release the anonymized dataset upon paper acceptance.

\begin{table*}[ht]
\centering
\caption{Definition and prevalence of cyberbullying roles used in this work. Count: the frequency (and percentage) of the 414 total sessions each role appears in.}
\begin{tabulary}{0.99\textwidth}{@{} l r L @{}}
\toprule
\textbf{Cyberbullying Role} & \multicolumn{1}c{\textbf{Count}} & \textbf{Definition} \\
\midrule
\multirow{2}{1in}{Bully} & \multirow[c]{2}{0.60in}{406 (98\%)} & Someone who attacks, harasses, humiliates, or threatens other people in the Instagram post. \\
\midrule
Bully Assistant & \multirow{1}{0.60in}{106 (26\%)} & Someone who sees bullying and begins to attack others in the comments. \\
\midrule
Aggressive Victim & \multirow{1}{0.60in}{118 (29\%)} & A person being attacked who responds aggressively to their attackers. \\
\midrule
\multirow{2}{1.5in}{Non-Aggressive Victim} & \multirow{2}{0.60in}{228 (55\%)} & A person being attacked who either ignores the attack or responds non-aggressively to it. \\
\midrule
\multirow{2}{*}{Aggressive Defender of Victim} & \multirow{2}{0.60in}{298 (72\%)} & A person who attempts to help someone being attacked, responding aggressively towards the attackers. \\
\midrule
\multirow{4}{*}{Non-Aggressive Defender of Victim} & \multirow{4}{0.75in}{\shortstack[l]{(A) 204 (49\%)\\[4pt](B) 203 (49\%)}\hspace{0.5em}} & A person who attempts to help someone being attacked through non-aggressive support. 
Exclusive to this rule, annotators had to choose a sub-type based on the intended target of the comment: (A) \textit{Directly Confronting a Bully} or (B) \textit{Supporting the Victim}.\\
\midrule
\multirow{2}{*}{Passive Bystander} & \multirow{2}{0.60in}{411 (99\%)} & A person who either has not seen any bullying or chooses to ignore any bullying-related comments they have seen. \\
\bottomrule
\end{tabulary}
\label{tab:bullying-roles}
\end{table*}
The dataset at the core of our analysis is an extended version of the Instagram cyberbullying dataset collected by~\citet{boulder_instragram_dataset}.
The original dataset contains sessions (posts) from 2013 to 2015, totaling 158,201 comments across 2,219 sessions.
Each session has an accompanying image and caption, a timestamp of when the session was posted, the session author's username, the number of likes (at time of collection) and other metadata.   
Because the original dataset included cyberbullying annotations at only the session-level, it was unsuitable for capturing user interactions at the \textit{comment-level}.
Thus, we selected a subset of 438 sessions (35,364 comments in total) based on the criteria that a majority of the five original annotators agreed that the session constituted cyberbullying.
With these 438 sessions, we used Amazon MTurk---a crowd-sourcing platform---to hire experienced \textit{Master} annotators to label and categorize all 35,364 comments.
All personally identifiable information was anonymized prior to being presented to the MTurk workers. 
Every comment was annotated by five independent annotators to ensure consistency.
In addition, the deployed survey had built-in attention checks to improve the annotation reliability.
Failing the attention checks lead to the disposal of the corresponding annotations.

In the survey, annotators were presented with a near replica of the original Instagram session page, where each comment had drop-down menus for annotating. 
When annotating a comment, the annotators were required to identify, (1) if the comment constituted bullying, (2) the comment author's  role, (3) the severity of the bullying, and (4) the topics of the bullying (\eg, gender identity, race, social status).
Severity is measured at three levels: \textit{mild}, \textit{moderate}, and \textit{severe}. 
The cyberbullying roles and their definitions are present in Tab.~\ref{tab:bullying-roles}.
We present these roles in title case throughout the manuscript. 

Crucial to our analyses, annotators were prompted at the end of the survey to pick the overall \textit{Main Victim} of the session. 
This session-level annotation determines who, after considering all comments in the session, is the primarily-impacted victim.
The choices were, (a) the user who created the post, designated OP for Original Poster, (b) people depicted in the picture, designated Picture, (c) the participants in the comments, designated Participants, or (d) Other.

\begin{table}[tb]
    \centering
    \caption{Example of comment annotations for a session. Each comment is annotated with a role and a severity score. 
    The following comments are used to construct the session graph in Fig~\ref{fig:session-digraph}. 
    The annotated Main Victim is the \textit{Poster}. 
    Users can take on different roles as comment threads progresses, \eg, $u_3$ is an Aggressive Defender, an Aggressive Victim, and a Non-Aggressive Victim within the same session. 
    Sequence: comment sequence id ordered by comment creation timestamp, Severity: average comment severity score across the majority annotators.
    }
    \pgfplotstableread[
        col sep=comma, trim cells,
    ]{./data-v2/session_comments.txt}\comments

        \pgfplotstabletypeset[
            trim cells=true,
            columns={comment_id,user,role,severity}, 
            columns/comment_id/.style={
                column name=\textbf{Sequence}, 
                fixed, column type={@{} c},
            },
            columns/user/.style={
                column name=\textbf{User},
                string type, 
                column type=c,
            },
            columns/role/.style={
                column name=\textbf{Role}, 
                string type, 
                column type=l,
            },
            columns/severity/.style={
                column name=\textbf{Severity},
                column type={r @{}},
                fixed, fixed zerofill, precision=2, 
            },
            every head row/.style={before row=\toprule,after row=\midrule},
            every last row/.style={after row=\bottomrule},
        ]\comments
    \label{tab:annotated-comments}
\end{table}

\BfPara{Preprocessing}
As noted earlier, each comment has five independent annotations. 
To arrive at aggregated role labels, we opt for a two-step approach. 
We first segregate annotators into two groups, based on the boolean option `is bullying' and `is not bullying'. 
We then take the annotators who are in the majority and disregard the minority annotators.
For example, if three annotators said `is bullying', but two annotators said `is not bullying', we discard the annotations from those who said not bullying. 

We then take another majority vote to decide the commenter's role.
It is possible for each of the remaining annotators to select a different role, hence, the need for a role tie-breaking heuristic. 
Of the 35,364 comments, 14,117 comments (40\%) have ties, mostly between two and three roles.
For cyberbullying identified comments, the role heuristic chooses a roles with the preference: \textit{Aggressive Defender} \textgreater \textit{ Aggressive Victim} \textgreater \textit{ Bully Assistant} \textgreater \textit{ Bully}.
For non-cyberbullying identified comments, the heuristic chooses a role with the preference: \textit{Non-Aggressive Defender: Support of the Victim} \textgreater \textit{ Non-Aggressive Defender: Direct to the Bully} \textgreater \textit{ Non-Aggressive Victim} \textgreater \textit{ Passive Bystander}. 
Notice that the heuristic is designed to prefer Defenders and Victims over Bullies and Passive Bystanders, as the latter two are the most common in the dataset.
In addition, after arriving at the final consensus label for each comment, we discard all comments with the Passive Bystander role. 
This role communicates non-participation in  cyberbullying sessions, \ie, the Passive Bystander comments do not actively shape the discourse.
Though, it can be argued that the mere presence of bystanders can influence the actions of other users. 
Despite that recognition, we opted to omit them from the present analyses due to the challenge of accurately measuring their influence.

To quantify a comment's annotated severity numerically, we use the following conversion scale.   
Non-cyberbullying and Mild annotations map to 1, Moderate maps to 2, and Severe to 3. 
Then we take the arithmetic mean across the majority annotators. 
Across all sessions, the average comment severity is 1.19.

Finally, to reach a consensus for the Main Victim of the session, first the Picture and the OP annotations are combined into a single category, \textit{Poster}, and then the overall victim is determined through another majority vote. 
In the 36 (8.2\%) sessions that contain a tie in their Main Victim majority, Poster is given preference over Participants. 
Because modeling the interactions between bullies and the victims is at the heart of this analysis, we subsequently filtered 24 sessions where the vote determined the Main Victim to be Other. 
Tab.~\ref{tab:annotated-comments} provides an illustration of the comment-level annotations of a session after applying the preprocessing step.
From this point forward in the paper, the analysis focuses on the 414 sessions containing valuable comment-level labels.

\section{Methods \& Results}
In this section, we describe our proposed frameworks and formalisms that substantiate our key contributions. 
The source code available as an anonymized repository\footnote{\url{https://anonymous.4open.science/r/cyberbullying-motifs/}}. 

\subsection{Session Graph}
We construct user interaction networks based on the processed comment-level annotations. 
These networks are henceforth referred to as \textit{Session Graphs}. Tab.~\ref{tab:session_stats_v2} summarizes key statistics across all the  constructed Session Graphs.  

\BfPara{Formalism} 
A Session Graph $\mathcal{G}$ is a directed graph represented by a 4-tuple  $\mathcal{G} = \langle V, E, \kappa, \rho \rangle$, where $V$ is a set of nodes; $E \subseteq V \times V$ is the set of directed edges; $\kappa: E \mapsto \mathbb{R}^{+}$ is a function assigning weights to edges; and 
$\rho: V \mapsto L$ is a function assigning nodes different labels, \ie, cyberbullying roles.
$L$ is the set of all possible cyberbullying roles (see Tab.~\ref{tab:bullying-roles} for definitions).
Edges have a default weight of one and the weights additively accumulate over repeated interactions between the same pair of nodes. 
An example Session Graph can be found in Fig.~\ref{fig:session-digraph} with 9 nodes and 18 edges. 
Roles are differentiated by both node shape and color.

For consistency and convenience, we define three special subsets of nodes that convey related roles.
Let $\mathcal{B}$ represent the combined set of Bully and Bully Assistant nodes.
Let $\mathcal{V}$ represent the combined set of Main Victim, Aggressive Victim, and Non-Aggressive Victim nodes.
And finally, let $\mathcal{D}$ represent the aggregated set of Aggressive and Non-Aggressive Defenders. Formally, we define the sets as:
\begin{align*}
    \begin{split}
    \mathcal{B} &= \{v \in V \mid \rho(v) \in \{\text{Bully, Bully Asst}\} \} \\
    \mathcal{V} &= \{v \in V \mid \rho(v) \in \{\text{Main Victim, Agg Victim, Non-Agg Victim}\}\}\\ 
    \mathcal{D} &= \{v \in V \mid \rho(v) \in \{\text{Agg Def, Non-Agg Def}\}\}
    \end{split}
\end{align*}

We use the terms \textit{Victims} and \textit{Bullies} interchangeably with $\mathcal{V}$ and $\mathcal{B}$, respectively.
We now detail how Session Graphs are constructed from the  annotated comment data. 



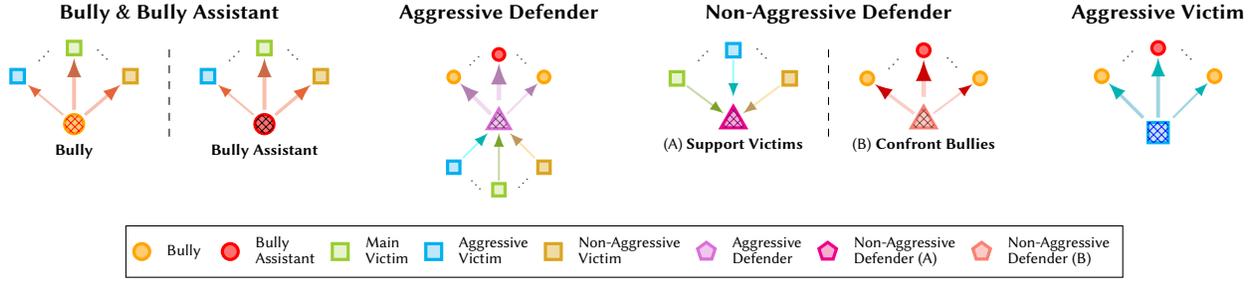
\begin{figure*}[tb]
    \centering
{
\begin{tikzpicture}
    \begin{groupplot}[
        group style={
            rows=1,
            columns=4,
            vertical sep=10pt,
            horizontal sep=50pt,
        },
        axis lines=none,   
        axis line style={draw opacity=0},
        tick style={draw=none},
        tick label style={font=\tiny},
        anchor=origin,
        clip=false,
        disabledatascaling,
        xmin=-1.75, xmax=1.75,
        ymin=-1.75, ymax=1.75,
        x=0.75cm, y=0.75cm,   
         title style={
            align=center, font=\small,
            yshift=-3pt, 
        },
    ]
  
    \nextgroupplot[
        title={\textbf{Bully} \textbf{\&} \textbf{Bully Assistant}},
        title style={yshift=0.75em, align=center},
    ]
        \begin{scope} [every node/.style={scale=0.6}, scale=1.0, shift={(-4em, 1em)}]
            \node[bully, scale=1.2, label=below:{\large \textbf{Bully}}, postaction={very thick, pattern=crosshatch, pattern color=Red}] (bul) at (0, 0.15) {};
            
            \node[aggressive_victim, scale=0.80] (av) at (-1, 1) {};
            \node[main_victim, scale=0.80] (mv) at (0, 1.5) {};
            \node[non_aggressive_victim, scale=0.80] (nav) at (1, 1) {};

            \node[textnode, rotate=45, font=\bfseries] at (-0.5, 1.35) {$\cdots$};
            \node[textnode, rotate=-45, font=\bfseries] at (0.5, 1.35) {$\cdots$};
            
            \draw[bully2main_vic, line width=1.5] (bul) edge (mv);
            \draw[bully2agg_victim] (bul) edge[bend left=0] (av);
            \draw[bully2agg_victim, line width=1.25] (bul) edge[bend right=0] (nav);
        \end{scope}

        \draw[dashed] (0, 0.35) -- (0, 2);
        
        \begin{scope} [every node/.style={scale=0.6}, scale=1.0, shift={(4em, 1em)}]
            \node[bully_asst, scale=1.2, label=below:{\large \textbf{Bully Assistant}}, postaction={very thick, pattern=crosshatch, pattern color=Black}] (bul) at (0, 0.15) {};
            \node[aggressive_victim, scale=0.80] (av) at (-1, 1) {};
            \node[main_victim, scale=0.80] (mv) at (0, 1.5) {};
            \node[non_aggressive_victim, scale=0.80] (nav) at (1, 1) {};

            \node[textnode, rotate=45, font=\bfseries] at (-0.5, 1.35) {$\cdots$};
            \node[textnode, rotate=-45, font=\bfseries] at (0.5, 1.35) {$\cdots$};
            
            \draw[bully2main_vic, line width=1.5] (bul) edge (mv);
            \draw[bully2agg_victim] (bul) edge[bend left=0] (av);
            \draw[bully2agg_victim, line width=1.25] (bul) edge[bend right=0] (nav);
    
        \end{scope}

        \nextgroupplot[
            title={\textbf{Aggressive Defender}},
            title style={yshift=0.75em, align=center},
        ]
        \begin{scope} [every node/.style={scale=0.55}, scale=0.80, shift={(0em, 1.80em)}]
            \node[aggressive_defender, scale=1.2, postaction={very thick, pattern=crosshatch, pattern color=Orchid!50!black}] (ag) at (0, 0) {};
            
            \node[aggressive_victim, scale=0.80] (av) at (-1, -1) {};
            \node[main_victim, scale=0.80] (mv) at (0, -1.5) {};
            \node[non_aggressive_victim, scale=0.80] (nav) at (1, -1) {};

            \node[bully, scale=0.80] (bul1) at (-1, 1) {};
            \node[bully, scale=0.80] (bul2) at (1, 1) {};
            \node[bully_asst, scale=0.80] (bul3) at (0, 1.5) {};

            \node[textnode, rotate=45, font=\bfseries] at (-0.5, 1.35) {$\cdots$};
            \node[textnode, rotate=-45, font=\bfseries] at (0.5, 1.35) {$\cdots$};

            \draw[main_vic2agg_def] (mv) edge (ag);
            \draw[agg_vic2agg_def] (av) edge[bend left=0] (ag);
            \draw[non_agg_vic2agg_def] (nav) edge[bend right=0] (ag);

            \draw[agg_def2bully, line width=1.5] (ag) edge (bul1);
            \draw[agg_def2bully] (ag) edge (bul2);
            \draw[agg_def2bully, line width=1.75] (ag) edge (bul3);

            \node[textnode, rotate=135, font=\bfseries] at (-0.5, -1.35) {$\cdots$};
            \node[textnode, rotate=-135, font=\bfseries] at (0.5, -1.35) {$\cdots$};
        \end{scope}

        \nextgroupplot[
            title={\textbf{Non-Aggressive Defender}},
            title style={yshift=0.75em, align=center},
        ]
        
        \draw[dashed] (0, 0.35) -- (0, 2);

        
        \begin{scope}[shift={(-4em,1.5em)}, every node/.style={scale=0.60}]
            \node[non_agg_defender_victim, scale=1.2, label={below:{\large (A) \textbf{Support Victims}}}, postaction={very thick, pattern=crosshatch, pattern color=magenta!50!black}] (nadV) at (0, 0) {};
            \node[main_victim, scale=0.80] (mv) at (-1, 0.75) {};
            \node[aggressive_victim, scale=0.80] (av) at (0, 1.25) {};
            \node[non_aggressive_victim, scale=0.80] (nav) at (1, 0.75) {};

            \draw[main_vic2agg_def] (mv) edge (nadV);
            \draw[agg_vic2agg_def] (av) edge (nadV);
            \draw[non_agg_vic2agg_def] (nav) edge (nadV);

            \node[textnode, rotate=45, font=\bfseries] at (-0.5, 1.15) {$\cdots$};
            \node[textnode, rotate=-45, font=\bfseries] at (0.5, 1.15) {$\cdots$};

        \end{scope}

        \begin{scope}[shift={(4em, 1.5em)}, every node/.style={scale=0.60}]
            \node[non_agg_defender_bully, scale=1.2, label={below:{\large (B) \textbf{Confront Bullies}}}, postaction={very thick, pattern=crosshatch, pattern color=Salmon!50!black}] (nadB) at (0, 0) {};
            \node[bully, scale=0.80] (bul1) at (-1, 0.75) {};
            \node[bully_asst, scale=0.80] (bul2) at (0, 1.25) {};
            \node[bully, scale=0.8] (bul3) at (1, 0.75) {};

            \draw[non_agg_def2bully, line width=1.25] (nadB) edge (bul1);
            \draw[non_agg_def2bully, line width=1.75] (nadB) edge (bul2);
            \draw[non_agg_def2bully] (nadB) edge (bul3);
            
            \node[textnode, rotate=45, font=\bfseries] at (-0.5, 1) {$\cdots$};
            \node[textnode, rotate=-45, font=\bfseries] at (0.5, 1) {$\cdots$};
        \end{scope}

         \nextgroupplot[
            title={\textbf{Aggressive Victim}},
            title style={yshift=0.75em, align=center},
        ]
            \begin{scope} [every node/.style={scale=0.6}, scale=1.0, shift={(0em, 1em)}]

                \node[aggressive_victim, scale=1.2, postaction={very thick, pattern=crosshatch, pattern color=Blue}] (av) at (0, 0) {};
                \node[bully, scale=0.80] (bul1) at (-1, 1) {};
                \node[bully_asst, scale=0.80] (bul2) at (0, 1.5) {};
                \node[bully, scale=0.8] (bul3) at (1, 1) {};

                \node[textnode, rotate=45, font=\bfseries] at (-0.5, 1.35) {$\cdots$};
                \node[textnode, rotate=-45, font=\bfseries] at (0.5, 1.35) {$\cdots$};

                \draw[agg_vic2bully, line width=1.25] (av) edge (bul1);
                \draw[agg_vic2bully, line width=1.45] (av) edge 
                (bul2);
                \draw[agg_vic2bully] (av) edge (bul3);  
            \end{scope}
    \end{groupplot}

    \begin{scope}[shift={(37em, -6em)}]
        \begin{customlegend}[ 
            legend columns=0,
            legend style={
            draw=black,
            fill=none,
            font=\scriptsize,
            column sep=0.5em,
            cells={align=left},
          },
          legend cell align={left},
          legend entries={
            {Bully~~},
            {Bully\\[-1pt]Assistant~~},
            {Main\\[-1pt]Victim~~},
            {Aggressive\\[-1pt]Victim~~~~},
            {Non-Aggressive\\[-1pt]Victim~~~},
            {Aggressive~~\\[-1pt]Defender~~},
            {Non-Aggressive~~\\[-1pt]Defender (A)~~},
            {Non-Aggressive\\[-1pt]Defender (B)~~},
        }
          ]
            \addlegendimage{bully, scale=2, mark options={scale=1.5}, only marks}
            \addlegendimage{bully_asst, scale=2, mark options={scale=1.5}, only marks}
            \addlegendimage{main_victim,  scale=2, mark options={mark=square*, scale=1.5}, only marks}
            
            \addlegendimage{aggressive_victim,  scale=2, mark options={mark=square*, scale=1.5}, only marks}
            \addlegendimage{non_aggressive_victim,  scale=2, mark options={mark=square*, scale=1.5}, only marks}

            \addlegendimage{aggressive_defender,  scale=2, mark options={mark=pentagon*, scale=1.75}, only marks}
            \addlegendimage{non_agg_defender_victim,  scale=2, mark options={mark=pentagon*, scale=1.75}, only marks}
            \addlegendimage{non_agg_defender_bully,  scale=2, mark options={mark=pentagon*, scale=1.75}, only marks}
            
            
        \end{customlegend}
    \end{scope}
\end{tikzpicture}
}
    \caption{Schematic description of the different edges induced by a new comment during the Session Graph creation process. The focal node in each panel is shaded. The Main Victim and Non-Aggressive Victim nodes do not result in the addition of new edges and are therefore absent in this figure.}
    \label{fig:edge-rules}
\end{figure*}

\BfPara{Graph Construction}
Consider a session $S$ with $k$ annotated comments where each comment $C$ has a timestamp $t$, the commenter's username $u$, the commenter's role $r$, and a severity score $w$.
By default, each Session Graph $\mathcal{G}$ starts with a single node representing the Main Victim (either the OP, Picture, or Participants).
Subsequent nodes are distinct tuples of the commenters' usernames and roles. 
Comments are processed sequentially, starting from the oldest to the newest, triggering the following topological changes in $\mathcal{G}$. 

\begin{enumerate}
    \item Add a new node in $\mathcal{G}$ corresponding to the pair of username and role values $(u, r)$ only if the current comment is the pair's first appearance in the session. Note, this  allows for multiple appearances of the same user in $\mathcal{G}$, provided they embody distinct roles in other comments. 
    
    \item Add new edges between the node $(u, r)$ and \textit{already existing} nodes following the guidelines described below. Instead of creating multiple edges between an already connected node pair, we simply increase the weight of the existing edge by the comment's severity score $w$. This is a similar strategy to what~\citet{kao_understanding_2019} proposed to track multiple interactions between users. This construction reflects the backwards looking perspective a user has when authoring a comment, wherein they respond to participants who have already written comments. 
\end{enumerate}
 
The guidelines for establishing edges from the current node ($\circ$) are presented below and are also illustrated in Fig.~\ref{fig:edge-rules}.

\begin{itemize}
    \item Bully and Bully Assistants attack the Victims. 
    We add separate directed edges of weight $w$, \ie, $\circ$ \arr $\mathcal{V}$.
    
    \item Aggressive Defenders simultaneously pacify the Victims and attack the Bullies.
    We establish two sets of edges, all with the same edge weight $w$. 
    The first set originate from the Victims and terminate at the current node, \ie, $\mathcal{V}$ \arr $\circ$.
    The second set of edges originate from the current node and are targeted towards the Bullies, \ie, $\circ$ \arr $\mathcal{B}$. 
    
    \item Non-Aggressive Defenders are unique in that they have a subtype. Either they confront the Bullies (Type A) or they support the Victims (Type B).
    For Type A, we add directed edges, $\circ$ \arr $\mathcal{B}$, with weight $w = 1$.
    Whereas Type B corresponds to directed edges, $\mathcal{V}$ \arr $\circ$, also with $w = 1$.
    
    \item Aggressive Victims confront the Bullies. 
    This is represented by separate directed edges of weight $w$ resembling \protect{$\circ$ \arr $\mathcal{B}$}.
    
    \item Non-Aggressive Victim and Main Victim nodes do not automatically establish any edges of their own. 
    They only receive edges as the result of actions taken by other nodes as they are introduced into the graph.
\end{itemize}

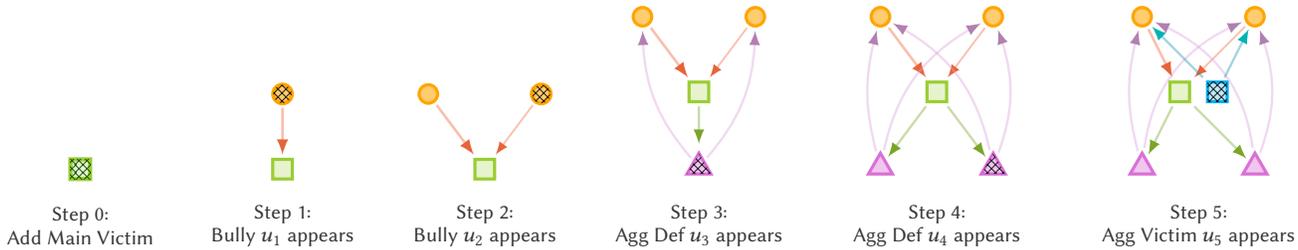
\begin{figure*}[tb]
    \centering
    \pgfdeclarelayer{background}
\pgfdeclarelayer{foreground}
\pgfsetlayers{background,main,foreground}

{
\begin{tikzpicture}[every node/.append style={scale=0.70}]
    \begin{scope}[shift={(-1em, 0em)}]   
        \node[textnode, scale=1.6] at (0, -1.75) {Step 0:\\Add Main Victim};
        
        \node[main_victim, ultra thick, postaction={very thick, pattern=crosshatch, pattern color=Green!50!black}] at (0, -1) (mv1) {};
    \end{scope}

    \begin{scope}[shift={(7.5em, 0em)}] 
        \node[textnode, scale=1.6] at (0, -1.75) {Step 1:\\Bully $u_1$ appears};
        
        \node[main_victim] at (0, -1) (mv) {};
        \node[bully, ultra thick, postaction={very thick, pattern=crosshatch, pattern color=black}] at (0, 0) (b1) {};

        \draw[bully2vic, line width=1] (b1) edge (mv);
    \end{scope}

    \begin{scope}[shift={(16em, 0em)}] 
        \node[textnode, scale=1.6] at (0, -1.75) {Step 2:\\Bully $u_2$ appears};
        
        \node[main_victim] at (0, -1) (mv) {};
        \node[bully] at (-0.75, 0) (b1) {};
        \node[bully, ultra thick, postaction={very thick, pattern=crosshatch, pattern color=black}] at (0.75, 0) (b2) {};

        \draw[bully2vic, line width=1] (b1) edge (mv);
        \draw[bully2vic, line width=0.71] (b2) edge (mv);
    \end{scope}

    \begin{scope}[shift={(25em, 0em)}] 
        \node[textnode, scale=1.6] at (0, -1.75) {Step 3:\\Agg Def $u_3$ appears};

        \begin{scope}[shift={(0em, 3.25em)}]
            \node[main_victim] at (0, -1) (mv) {};
            \node[bully] at (-0.75, 0) (b1) {};
            \node[bully] at (0.75, 0) (b2) {};
    
            \node[aggressive_defender, postaction={very thick, pattern=crosshatch, pattern color=black}] at (0, -2) (ad1) {};
    
            \draw[bully2vic, line width=1] (b1) edge (mv);
            \draw[bully2vic, line width=0.71] (b2) edge (mv);
            
            \draw[agg_def2bully, line width=0.89] (ad1) edge[bend left=20] (b1);  
            \draw[agg_def2bully, line width=0.89] (ad1) edge[bend right=20] (b2);  

            \draw[main_vic2agg_def, line width=0.89] (mv) edge (ad1);  
            
        \end{scope}
    \end{scope}

    \begin{scope}[shift={(35em, 0em)}] 
        \node[textnode, scale=1.6] at (0, -1.75) {Step 4:\\Agg Def $u_4$ appears};

        \begin{scope}[shift={(0em, 3.25em)}]
            \node[main_victim] at (0, -1) (mv) {};
            \node[bully] at (-0.75, 0) (b1) {};
            \node[bully] at (0.75, 0) (b2) {};

            \node[aggressive_defender] at (-0.75, -2) (ad1) {};
            \node[aggressive_defender, postaction={very thick, pattern=crosshatch, pattern color=black}] at (0.75, -2) (ad2) {};
    
            \draw[bully2vic, line width=1] (b1) edge (mv);
            \draw[bully2vic, line width=0.71] (b2) edge (mv);
            
            \draw[agg_def2bully, line width=0.89] (ad1) edge[bend left=15] (b1);  
            \draw[agg_def2bully, line width=0.89] (ad1) edge[bend left=25] (b2);  
            \draw[main_vic2agg_def, line width=0.89] (mv) edge (ad1);  

            \draw[agg_def2bully, line width=0.89] (ad2) edge[bend right=25] (b1);  
            \draw[agg_def2bully, line width=0.89] (ad2) edge[bend right=20] (b2);  
            \draw[main_vic2agg_def, line width=0.89] (mv) edge (ad2);  
        \end{scope}
    \end{scope}

    \begin{scope}[shift={(46em, 0em)}] 
        \node[textnode, scale=1.6] at (0, -1.75) {Step 5:\\Agg Victim $u_5$ appears};
        
        \begin{scope}[shift={(0em, 3.25em)}]
            \node[main_victim] at (-0.25, -1) (mv) {};
            \node[aggressive_victim, postaction={very thick, pattern=crosshatch, pattern color=black}] at (0.25, -1) (av1) {};
            
            \node[bully] at (-0.75, 0) (b1) {};
            \node[bully] at (0.75, 0) (b2) {};

            \node[aggressive_defender] at (-0.75, -2) (ad1) {};
            \node[aggressive_defender,] at (0.75, -2) (ad2) {};
    
            \draw[bully2vic, line width=1] (b1) edge (mv);
            \draw[bully2vic, line width=0.71] (b2) edge (mv);
            
            \draw[agg_def2bully, line width=0.89] (ad1) edge[bend left=15] (b1);  
            \draw[agg_def2bully, line width=0.89] (ad1) edge[bend left=30] (b2);  
            \draw[main_vic2agg_def, line width=0.89] (mv) edge (ad1);  

            \draw[agg_def2bully, line width=0.89] (ad2) edge[bend right=30] (b1);  
            \draw[agg_def2bully, line width=0.89] (ad2) edge[bend right=20] (b2);  
            \draw[main_vic2agg_def, line width=0.89] (mv) edge (ad2);  

            \draw[agg_vic2bully, line width=0.89] (av1) edge (b1);  
            \draw[agg_vic2bully, line width=0.89] (av1) edge (b2);  
        \end{scope}
    \end{scope}
\end{tikzpicture}
}
    \caption{Step-by-step construction of the Session Graph $\mathcal{G}$ based on the first five comments in Tab.~\ref{tab:annotated-comments}. The new node introduced in each step is shaded. Roles are abbreviated to save space.}
    \label{fig:step-by-step}
\end{figure*}

We illustrate the graph building process in Fig.~\ref{fig:step-by-step} by utilizing the first five comments found in Tab.~\ref{tab:annotated-comments}. After processing the remaining three comments, we arrive at the partially completed Session Graph shown in Fig.~\ref{fig:session-digraph}. Walking through the process, starting with comments one and two, as there are no bullies currently present in the graph, the only edges that are added are between the Main Victim node to the two Bullies (users $u_1$ and $u_2$) with weights 2.00 and 1.00, respectively. 
For the third comment, since $u_3$ is Aggressive Defender, this node will establish a directed edge with weight 1.60 to the Main Victim node and two directed edges with weight 1.60 to the Bullies. 
For comment four, since $u_4$ is also an Aggressive Defender, this node establishes a directed edge with weight 1.67 to the Main Victim and two directed edges with the same weight to the two Bully nodes. 
Finally, for comment five, as $u_5$ is an Aggressive Victim, it establishes two edges with the \textit{Bullies}, one with $u_1$ and the other with $u_2$, each of weight 1.60.

\subsection{Victim \& Bully Scores}
To both quantify and differentiate the activity of Victims and Bullies within a session, we leverage network metrics 
that focus their attention on the subsets of nodes $\mathcal{B}$ and $\mathcal{V}$ that we defined earlier.
For each subset, we define a numeric \textit{score} that is the average of the differences between the \textit{weighted out-} and \textit{weighted in-degrees} of each sets member nodes.
Specially, we define these scores as the \textit{Victim Score} and the \textit{Bully Score}: 
\begin{align*}
    \text{Victim Score} &= \frac{\sum_{v\in \mathcal{V}} \left( d_w^{\text{out}}(v) - d_w^{\text{in}}(v) \right)}{|\mathcal{V}|} \\
    \text{Bully Score} &= \frac{\sum_{v\in \mathcal{B}} \left( d_w^{\text{out}}(v) - d_w^{\text{in}}(v) \right)}{|\mathcal{B}|},
\end{align*}
where $d_w^{\text{out}}$ and $d_w^{\text{in}}$ represent the weighted in- and out-degree of a node in $\mathcal{G}$.

These scores, when considered in unison, communicate the balance of power present between the Victims and the Bullies of a session. 
A high Victim Score reflects that Victims receive outweighed support from Defenders, a lack of Bullies, or both, whereas a high Bully Score indicates that the bullying occurred without significant pushback from Defenders or Aggressive Victims. 

\begin{figure}[tb!]
    \centering
    \pgfplotstableread[col sep=comma]{./data-v2/session-stats.txt}\sessions 

\pgfplotstablecreatecol[
    create col/expr={
        (\thisrow{victim_score_weighted} < 0) ? nan : \thisrow{victim_score_weighted}
    }
]{victim_score_weighted_pos}{\sessions}

\pgfplotstablecreatecol[
    create col/expr={
        (\thisrow{victim_score_weighted} >= 0) ? nan : \thisrow{victim_score_weighted}
    }
]{victim_score_weighted_neg}{\sessions}

\pgfplotstablecreatecol[
    create col/expr={
        (\thisrow{bully_score_weighted} < 0) ? nan : \thisrow{bully_score_weighted}
    }
]{bully_score_weighted_pos}{\sessions}
\pgfplotstablecreatecol[
    create col/expr={
        (\thisrow{bully_score_weighted} >= 0) ? nan : \thisrow{bully_score_weighted}
    }
]{bully_score_weighted_neg}{\sessions}

\begin{tikzpicture}
    \begin{groupplot}[
        group style={
            rows=2,
            columns=1,
            xticklabels at=all,
            xlabels at=all,
            vertical sep=45pt,
        },
        area style,
        xtick=,%
        title style={
            yshift=-5pt,
        },
        xlabel style={
            font=\normalsize,
        },
        enlarge x limits=0.05,
        enlarge y limits=upper,
        ylabel style={
            yshift=-3pt,
            font=\normalsize,
        },
        x tick label style={font=\scriptsize},
        y tick label style={font=\scriptsize},
        ymajorgrids,
        grid style={dashed, gray!50, draw opacity=0.5},
        width=0.51\textwidth,
        height=140pt,
        axis x line=bottom,
        axis y line=left,
        ybar legend,
        legend columns=0,
        legend style={
            column sep=0.1em,
            draw=black,
            inner xsep=2.5pt,
            inner ysep=2pt,
            fill=white,
            font=\footnotesize,
            at={(11em, -19.25em)},
            anchor=south,
        },
        nodes near coords style={
            font=\scriptsize,
        },
        nodes near coords greater equal only={1},
    ]
        
        \nextgroupplot[
            ylabel={\# Sessions},
            xmin=-205, xmax=105,
            ymax=160,
            ytick distance=50,
            xlabel={Victim Score},
        ]
        \addplot[
            ybar interval,
            hist={
                bins=20,
                data=y,
                data min=-200,
                data max=0,
            },
            thick, draw=Red, fill=Red!50,
        ]
        table[
            y=victim_score_weighted_neg,
        ] {\sessions}; 

        \addplot[
            forget plot,
            ybar interval,
            hist={
                bins=20,
                data=y,
                data min=-200,
                data max=0,
                intervals=false,
            },
            draw=none, fill=none,
            nodes near coords={\pgfmathprintnumber[assume math mode=true]{\pgfplotspointmeta}},
            nodes near coords style={
                text=Red, 
                font=\scriptsize\sffamily,
                xshift=3.5pt,
            },
        ]
        table[
            y=victim_score_weighted_neg,
        ] {\sessions}; 

        \addplot[
            hist={
                bins=11,
                data=y,
                data min=0,
                data max=110,
            },
            thick, draw=Cerulean!70!black, fill=Cerulean!50,
            postaction={pattern=north west lines, pattern color=Cerulean!50!black},
            nodes near coords={\pgfmathprintnumber[assume math mode=true]{\pgfplotspointmeta}},
            nodes near coords style={
                text=Cerulean!50!black, 
                font=\scriptsize\sffamily,
                xshift=4.5pt,
            },
        ]
        table[
            y=victim_score_weighted_pos,
        ] {\sessions}; 

        \nextgroupplot[
            ylabel={\# Sessions},
            xmin=-80, xmax=155,
            ymax=230,
            xlabel={Bully Score},
        ]
        \addplot[
            ybar interval,
            hist={
                bins=6,
                data max=0,  
                data min=-60,
            },
            thick, draw=Red, fill=Red!50,
        ]
        table[
            y=bully_score_weighted_neg,
        ] {\sessions}; 

        \addplot[
            forget plot,
            ybar interval,
            hist={
                bins=6,
                data=y,
                data min=-60,
                data max=0,
                intervals=false,
            },
            draw=none, fill=none,
            nodes near coords={\pgfmathprintnumber[assume math mode=true]{\pgfplotspointmeta}},
            nodes near coords style={
                text=Red, 
                font=\scriptsize\sffamily,
                xshift=3pt,
            },
        ]
        table[
            y=bully_score_weighted_neg,
        ] {\sessions}; 

        \addplot[
            ybar interval,
            hist={
                bins=16,
                data min=0,
                data max=160,
            },
            thick, draw=Cerulean!70!black, fill=Cerulean!50,
            postaction={pattern=north west lines, pattern color=Cerulean!50!black},
        ]
        table[
            y=bully_score_weighted_pos,
        ] {\sessions}; 

        \addplot[
            forget plot,
            ybar interval,
            hist={
                bins=16,
                data min=0,
                data max=160,
                intervals=false,
            },
            draw=none, fill=none,
            nodes near coords={\pgfmathprintnumber[assume math mode=true]{\pgfplotspointmeta}},
            nodes near coords style={
                text=Cerulean!50!black, 
                font=\scriptsize\sffamily,
                xshift=3.5pt,
            },
        ]
        table[
            y=bully_score_weighted_pos,
        ] {\sessions};
    \end{groupplot}

    \node[text=Red, fill=white, font=\small, align=center] at (6em, 7.5em) {345 Sessions\\(84\%)};
    \node[text=Cerulean, fill=white, font=\small, align=center] at (20.5em, 7.5em) {69 Sessions\\(16\%)};

    \node[text=Red, fill=white, font=\small, align=center] at (3.2em, -8.25em) {141 Sessions\\(34\%)};
    \node[text=Cerulean, fill=white, font=\small, align=center] at (17em, -8.25em) {273 Sessions\\(66\%)};

    \node[text=black, font=\large\bfseries\sffamily] at (-1em, 11.5em) {A};
    \node[text=black, font=\large\bfseries\sffamily] at (-1em, -4em) {B};
\end{tikzpicture}
    \caption{Victim Score (panel A) and Bully Score (panel B) distributions for $N = 414$ sessions. 
    Positive scores are marked in blue, while the negative scores are in red. 
    Bin width in both panels is 10.
    Victim Scores are strongly skewed to the left, whereas the Bully Scores have a slight right skew. 
    }
    \label{fig:victim-score-dist}
\end{figure}
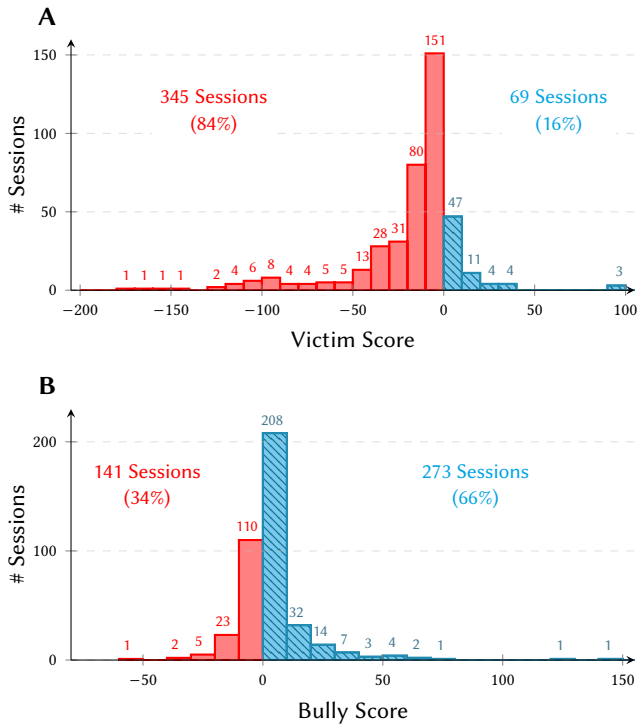

We now look at the trends for the scores across the dataset.
Fig.~\ref{fig:victim-score-dist}(A) presents the distribution of the Victim Score and 
Tab.~\ref{tab:session_stats_v2} offers the median of said distribution at -8.98. 
The Victim Score distribution is visibly left-skewed, with 345 (84\%) (highlighted in red) of the scores being negative.
Only 69 (16\%) (highlighted in blue) are positive. 
Similarly, Fig.~\ref{fig:victim-score-dist}(B) presents the distribution of the Bully Score. 
Per Tab.~\ref{tab:session_stats_v2}, the median Bully Score is 1.33.
This distribution has a slight right skew, with 141 (34\%) (shown in red) negative scores compared to 273 (66\%) (shown in blue) positive scores. 
The Victim Score distribution shows considerably greater variability, as evidenced by its long, drawn-out tail of negative scores. 
This is also reflected in the width of the confidence intervals for the mean (Tab.~\ref{tab:session_stats_v2}). 
Considering the average number of Victim nodes is 3.68, this suggests Victim nodes have substantially more incoming edges from Bullies, as opposed to outgoing edges from the presence of Defenders.
The distribution of the Bully Scores, on the other hand, contains less variability.

\begin{table}[tb]
    \caption{Session Graph statistics for 414 cyberbullying sessions. The Mean CI columns represent the 95\% confidence intervals around the mean. 
    On average, graphs typically have 24 nodes and 85 edges, with Bullies outnumbering the Victims by a factor of 4. 
    Victims display a wide weight disparity across their incoming and outgoing links, manifesting in their highly negative score.
    Bullies have a more balanced incoming and outgoing connection intensities.
    }
    {\small 
    \begin{tabular}{@{}ll rrrr@{}}
        \toprule
                                 &   & \multirow{2}[4]{*}[5pt]{\textbf{Median}} & \multirow{2}[4]{*}[5pt]{\textbf{Mean}} & \multicolumn{2}{c}{\textbf{Mean CI}} \\ \cmidrule(l){5-6} 
                                 &     &  &                    & \multicolumn{1}{c}{\textbf{Lower}}      & \textbf{Upper}     \\ \midrule
        \textbf{Nodes}                    &  & 17 & 23.65 & 21.74 & 25.74 \\ \midrule
        \textbf{Edges}                    &  & 41 & 85.14 & 74.49 & 99.04 \\ \midrule 
        \multirow{4}{*}{\textbf{Victims}} & Count & 2 & 3.68 & 3.33 & 4.11 \\[2pt] 
                                 & Weighted Out-degree    & 4.44  & 8.53 & 7.45 & 9.97 \\
                                 & Weighted In-degree     & 14.72 & 26.04 & 23.31 & 29.25 \\[2pt]
                                 & Score                  & $-$8.98 & $-$17.51 & $-$20.79 & $-$14.72  \\ 
                                 \midrule
        \multirow{4}{*}{\textbf{Bullies}} & Count         & 8 & 13.19 & 11.86 & 14.90 \\[2pt]
                                 & Weighted Out-degree    & 3.32 & 9.03  & 7.74  & 10.83 \\ 
                                 & Weighted In-degree     & 2.63 & 5.39  & 4.74  & 6.14  \\[2pt]
                                 & Score                  & 1.33 & 3.64  & 2.32  & 5.35  \\
                                 \bottomrule
    \end{tabular}
    }
    \label{tab:session_stats_v2}
\end{table}

\begin{figure*}[htb!]
    \centering
    \input{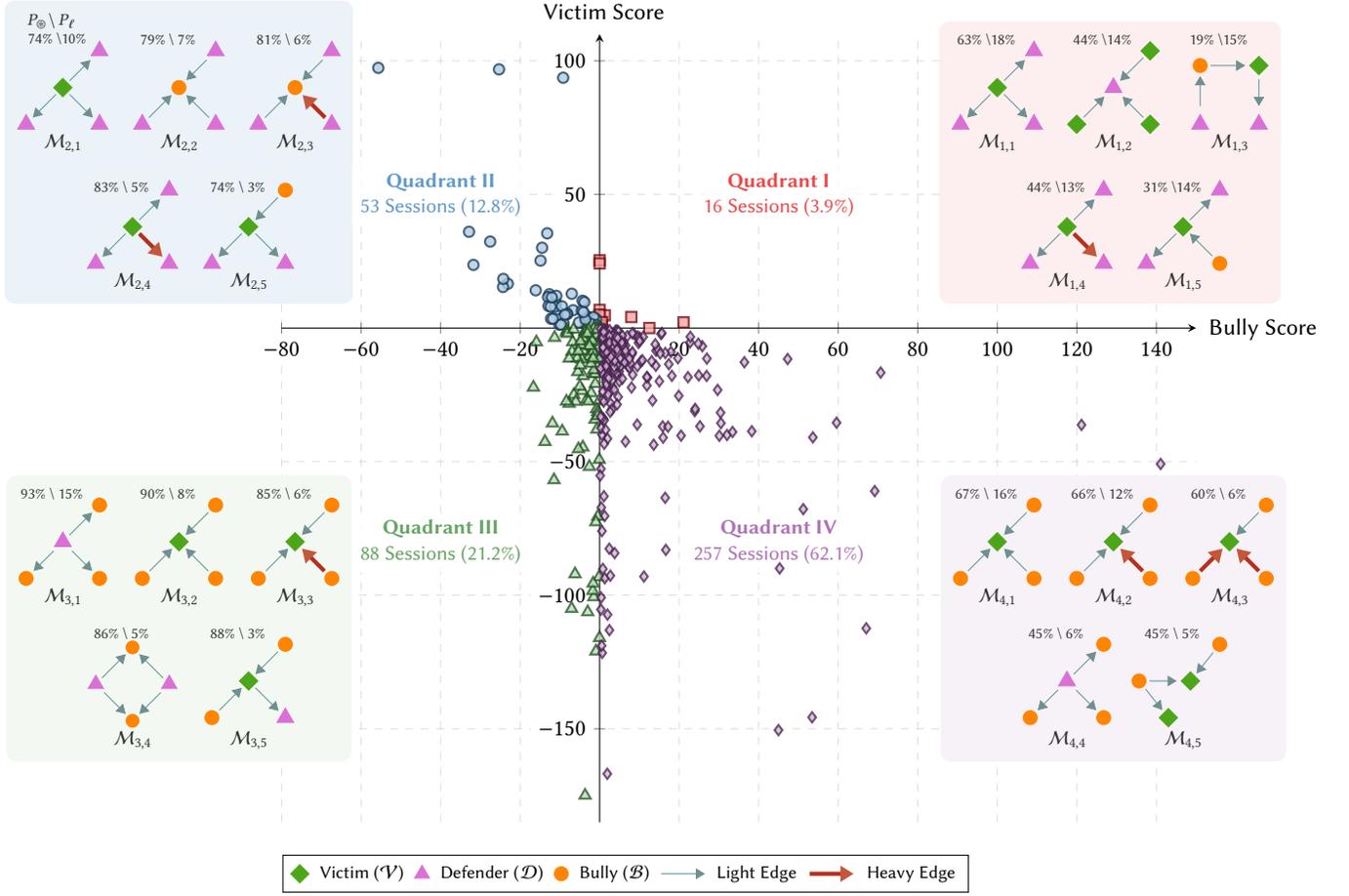}
    \caption{Joint distribution plot of the Bully and Victim Scores for 414 sessions. 
    The main plot is divided into four quadrants.
    Quadrant I (red squares) represents evenly matched sessions; In Quadrant II (blue circles) Victims and Defenders Dominate; Bullies face pushback in Quadrant III (green triangles), and Bullies dominate in Quadrant IV (purple diamonds). 
    Within each quadrant, we enumerate the top 5 motifs based on a combined Local and Global Prevalence ranking. 
    Motifs are indexed $\mathcal{M}_{i, j}$, where $i \in \{ 1, 2, 3, 4 \}$ is the quadrant and $j \in \{ 1, \ldots, 5 \}$ is the within-quadrant rank. 
    Node shapes denote roles: Victim (green diamond), Defender (purple triangle), and Bully (orange circle). 
    Directed, weighted edges indicate the flow and intensity of interactions within each motif.
    }
    \label{fig:joint-plot}
\end{figure*}

Fig.~\ref{fig:joint-plot} presents the joint distribution of the  Bully and Victim scores.
Each quadrant represents one of four possible combinations of the different signs assumed by the Bully Score and the Victim Score, respectively. 
Therefore, these quadrants communicate different outcomes in the bully-victim dynamic.
For instance, the session depicted in Fig.~\ref{fig:session-digraph} would be located in Quadrant IV where Bullies dominate the discourse.
Both Scores are simultaneously positive in the 16 sessions (3.9\%) present in Quadrant I.
These sessions represent an evenly matched scenario where the amount of aggression from the Bullies is displaced by nearly equal support by Defenders and Aggressive Victims. 
Quadrant II, with 53 sessions (12.8\%), demonstrates when Victims and Defenders dominate the interaction. 
They either vastly outnumber the Bullies or overwhelm them by flooding support to the Victims. 
In Quadrant III, which contains 88 sessions (21.2\%), Bullies face pushback from the Aggressive Victims and Defenders.
The simultaneously negative Victim and Bully scores suggest repeated back and forth between the two factions, with the Bullies maintaining a slight advantage. 
Finally, Quadrant IV is where most sessions reside (257 sessions, 62.1\%). 
It is also the quadrant where Bullies clearly dominate the interactions.
Despite the wide range of scores, we find a dense concentration of sessions around the origin. This implies that in many sessions, the positive and the negative interactions mutually neutralize each other.

\subsection{Motifs} 
As indicated earlier, we seek to understand the prevalence of certain interaction patterns between nodes of different roles.
We achieve this by tracking distributions of substructures called \textit{motifs} within the Session Graphs.
Motifs can be considered the building blocks that define a graph. Therefore, they also provide a mesoscale view of connectivity patterns in a larger graph.
For our purposes, we consider a motif $\mathcal{M}$ to be a connected, induced subgraph of a larger graph containing either three or four nodes.
We obtain motifs by using the iGraph library's implementation of the FANMOD algorithm on our Session Graphs $\mathcal{G}$ ~\cite{wernicke2006fanmod,igraph}.   
It is important to note that a specific motif $\mathcal{M}$ may appear multiple times within a graph, and similarly a graph may contain several distinct motifs.

\begin{figure}[tb]
    \centering
    \input{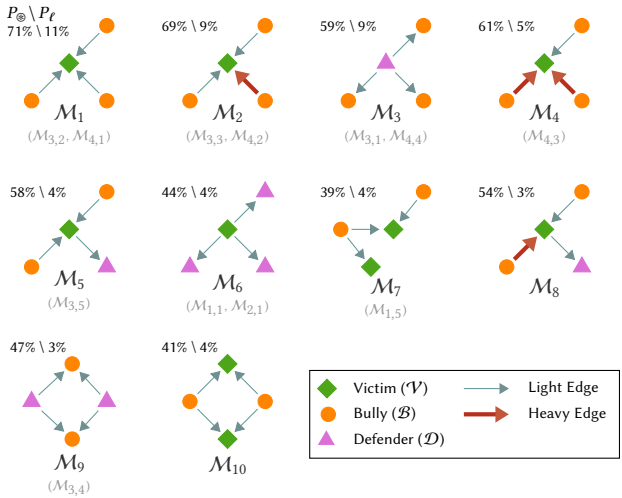}

    \caption{ 
        Overall top 10 motifs with high Global Prevalence ($P_\circledast$) and high Local Prevalence ($P_\ell$) scores. 
        Motifs are indexed $\mathcal{M}_{1}$ through $\mathcal{M}_{10}$ in ascending order of the rank product. 
        Most motifs also appear in Fig.~\ref{fig:joint-plot}, so we 
        also list those indices in gray.
        Each motif is annotated with its $P_\circledast$~\textbackslash~ $P_\ell$ scores. 
        Node shapes represent roles: Victim (green diamond), Defender (purple triangle), and Bully (orange circle). 
        Directed, weighted edges indicate the flow and intensity of interactions within each motif.
    }
    \label{fig:motifs-global}
\end{figure}

Our motif analysis strategy applies a few simplifications to the Session Graph structure. 
We aggregate roles corresponding to $\mathcal{V}$, $\mathcal{B}$, and $\mathcal{D}$ before running the motif search algorithm. 
So, for example, the Main Victim, Non-Aggressive Victims, and Aggressive Victims are mapped to a singular role \textit{Victim}. 
We also group the edges by weight into two buckets, \textit{light} (weight \textless 2) and \textit{heavy} (weight $\geqslant$ 2). 
These simplifications reduce the number of unique motif patterns to analyze, simplify the presentation of the most frequently occurring structures while preserving the most frequently occurring patterns, and let us differentiate between one-off or mild interactions and the repeated, more intense kinds. 
Figures~\ref{fig:joint-plot} and ~\ref{fig:motifs-global} provide examples of motifs that were found as part of the enumeration process. 


Session Graphs typically contain a wide variety of motifs.
Therefore, we rely on two complementary statistical measures to quantify the significance of particular motifs.
These measures are designed to capture a motif $\mathcal{M}$'s \textit{prevalence} or relative importance. 
The first measure, \textit{Local} Prevalence, is the prevalence of a motif localized to a specific session. While the second measure, \textit{Global} Prevalence, considers the prevalence of a motif across multiple sessions.
We provide their formal definitions below.

Let $\mathcal{S} = \{S_1, \cdots, S_m\}$ represent the set of $m$ sessions currently under consideration. 
We define a function $f(\mathcal{M}, S)$ that counts the frequency of occurrence of a given motif $\mathcal{M}$ in a session $S \in \mathcal{S}$.

\BfPara{Global Prevalence} $P_\circledast(\mathcal{M}, \mathcal{S})$ measures the proportion of sessions within $\mathcal{S}$ where the motif $\mathcal{M}$ appears at least once. 
\begin{align*}
    P_\circledast(\mathcal{M}, \mathcal{S}) &= \frac{\lvert\{ S \mid f(\mathcal{M}, S) > 0, S \in \mathcal{S} \}\rvert}{|\mathcal{S}|} 
\end{align*}
This measure captures how widespread a motif is across a set of sessions, regardless of how frequently it appears within any individual session. 
Even if a motif is rare within each session, a motif with high $P_\circledast$ score shapes the background dynamics through its frequent reappearance. 

\BfPara{Local Prevalence} $P_\ell(\mathcal{M}, \mathcal{S})$ is the average probability of observing the motif $\mathcal{M}$ within the sessions it appears in.
\begin{align*}
    P_\ell(\mathcal{M}, \mathcal{S}) &= \frac{1}{|\mathcal{S}|} \sum_{S \in \mathcal{S}} \left( \frac{f(\mathcal{M}, S)}{\sum_{m \in \mu(S)} f(m, S)} \right)
\end{align*}
The Local Prevalence calculation has the additional dependency on the set $\mu(S) = \{ \mathcal{M}_1, \cdots, \mathcal{M}_n \}$, the set of $n$ distinct motifs that appear in a given session $S$. 
The inner fraction measures the relative frequency of $\mathcal{M}$ compared to the rest of the motifs appearing in session $S$. 
The outer summation serves to compute the average. 

When a motif manifests in a session, Local Prevalence ($P_\ell$) is a indicator of its intensity. 
For example, motifs that appear in many sessions but at low frequencies would have a low $P_\ell$ score. 
In contrast, motifs that appear in fewer sessions overall but recur frequently within those sessions would have a high $P_\ell$ score.
By focusing our analysis on motifs with high $P_\ell$, we avoid overemphasizing those that are scarcely present and therefore do not meaningfully shape the sessions’ dynamics.

\BfPara{Motif Ranking Framework}
To identify the most prominent motifs, we apply a ranking framework based on the two measures $P_\circledast$ and $P_\ell$. 
Depending on the context, $\mathcal{S}$ is either set as the entire set of sessions, or the sessions within an individual quadrant found in Fig.~\ref{fig:joint-plot}.  
Motifs are first ranked separately by $P_\circledast$ and by $P_\ell$ using dense ranking, with the highest values assigned rank 1.
We then compute a combined rank as the product of these two ranks. 
This formulation emphasizes motifs with high Global and Local Prevalence scores since only motifs that perform strongly on both measures yield a low rank product. 
Finally, motifs are ordered by increasing rank product, with the lowest values indicating motifs that score strongly on both measures.
Fig.~\ref{fig:motifs-global} shows the top ranked motifs across the entire dataset while Fig.~\ref{fig:joint-plot} displays the top performers within each quadrant in separate boxes.
We will describe these results next.

\BfPara{Global Motif Patterns}
In Fig.~\ref{fig:motifs-global}, we show the overall top 10 motifs identified across all 414 sessions in the dataset. 
We set $\mathcal{S}$ to span all sessions while calculating the motif prevalence scores.
Following this process, three notable structural patterns emerge.
(1) Mild, repeated low-impact bullying. In this pattern, Bullies repeatedly attack the Victims.
This pattern is the most prevalent motif structure across all sessions, though with some variation introduced through the intensity of the edge weights.
In short, we find that repeated low-intensity or one-off attacks are more widespread than severe exchanges of hostility.
We can also consider this structure to be a defining characteristic of cyberbullying sessions.
(2) Intense bullying and heavy intervention occur intermittently.  
While heavy edges appear in several top-ranking motifs, they do not dominate the structures.
They tend to be localized, highlighting specific repeated, intense attacks within otherwise light interactions. 
This indicates that intense harassment is present, but not pervasive. 
(3) Defender involvement is common, but is of low-intensity. 
The presence of Defenders in high-ranking motifs shows that intervention is a regular feature of the sessions, but the edges they contribute are predominantly light.
This reflects quick attempts to counteract bullying, though Defenders do not engage in prolonged interventions.
Throughout the top motifs, bullying, victimization, and defending behaviors frequently co-occur. The edge weights reveal that these multi-actor structures are often held together by streams of light weight interactions. 

    
\BfPara{Quadrant-specific Motif Patterns}
Next, we zoom in on the top 5 motifs in each of the four quadrants in Fig.~\ref{fig:joint-plot}. 
Here, we set $\mathcal{S}$ to the sessions present in a given quadrant before computing the motif prevalence scores. 
For example, the Local Prevalence score is calculated only across the sessions in that quadrant. 
This means this value can differ significantly from the \textit{overall} Local Prevalence score, as the latter is calculated across all sessions.
We summarize our findings below.

Quadrant I is the smallest quadrant by far, with only 16 sessions. 
While the quadrant accounts for a tiny portion of the dataset, it still has influence. 
Notably, the $\mathcal{M}_{1,1}$ in Fig.~\ref{fig:joint-plot} that depicts a Victim reaching out to three Defenders for support with light edges also appears among the overall top 10 motifs ($\mathcal{M}_6$). 
The motifs $\mathcal{M}_{1,1}$ and $\mathcal{M}_{1,5}$ also appear in the top 5 for Quadrant II. 
The motifs in Quadrant I are characterized by Defender-Victim dynamics, with Defenders playing a central and structurally consistent role across the top five motifs. 
Another key feature of these motifs is the predominance of light interactions. Overall, these motifs represent situations in which Victims are shielded by multiple Defenders. 
This configuration reflects scenarios where several users step in to support a Victim, each contributing support or counter-attacking the Bullies. 
These patterns in unison reveal that in Quadrant I, cyberbullying interactions are defined less by aggression and more by low-intensity defensive behavior.

Quadrant II contains 53 sessions and is defined by strong Defender activity. 
Across the top motifs, Defenders consistently cluster around either a Victim or a Bully, intervening through light, low-intensity interactions. 
Heavy edges appear sporadically ($\mathcal{M}_{2,3}$ and $\mathcal{M}_{2,4}$ in Fig.~\ref{fig:joint-plot}), indicating isolated moments of escalation rather than sustained conflict. 
Overall, this quadrant reflects situations where Defenders frequently intervene, shaping the interaction pattern through repeated but brief supportive or punitive actions. 
The quadrant also contributes to the global landscape, with $\mathcal{M}_{2,1}$ appearing among the overall top 10 motifs ($\mathcal{M}_6$).

Quadrant III is the second largest quadrant with 88 sessions and is characterized by high Bully activity, with Bullies comprising the majority of nodes in the top motifs. 
All but one interaction is light, which reflects the low-intensity attacks directed at either a Victim or a Defender. 
Only one heavy edge appears in ($\mathcal{M}_{3,3}$ in Fig.~\ref{fig:joint-plot}), which shows that there is occasional escalation, but it is very rare. 
Across the motifs, Victims and Defenders are outnumbered by Bullies. 
In three of the five motifs, $\mathcal{M}_{3,2}$, $\mathcal{M}_{3,3}$, and $\mathcal{M}_{3,5}$, Bullies are actively targeting a Victim, reflecting persistent and coordinated bullying behavior. 
Some resistance does emerge, most notably in $\mathcal{M}_{3,1}$, where a single Defender confronts three Bullies. 
Although less common, the Defender-driven motifs does have influence, with a Global Prevalence of 93\% and a Local Prevalence of 15\%. 
Overall, Quadrant III captures environments characterized by widespread bullying activity alongside selective defensive intervention. 
The motifs identified here have strong overall influence with $\mathcal{M}_{3,1}$ through $\mathcal{M}_{3,5}$ appearing in the overall top 10 motifs, highlighting their importance across the dataset.

Bullies dominate in Quadrant IV. 
This is by far the largest quadrant, with 257 sessions (62\% of the dataset), and the most influential quadrant, shaping the motif landscape across the entire dataset. 
$\mathcal{M}_{4,1}$ through $\mathcal{M}_{4,4}$ are also the top four motifs overall ($\mathcal{M}_{1}$--$\mathcal{M}_{4}$ in Fig.~\ref{fig:motifs-global}), emphasizing the dominance of the structural patterns in this quadrant. 
In every top motif, Bullies appear at substantially higher frequencies than Defenders and Victims.  
This quadrant exhibits the greatest concentration of heavy edges among all quadrants, appearing in $\mathcal{M}_{4,2}$ and $\mathcal{M}_{4,3}$. 
Although motifs $\mathcal{M}_{4,1}$ through $\mathcal{M}_{4,3}$ share the same underlying structure of three Bullies targeting a single Victim, differences in edge weights reveal meaningful variation in the intensity of these interactions. 
The first motif, $\mathcal{M}_{4,1}$, contains only light edges; however, as the ranking progresses, heavy edges begin to appear. 
This progression provides a clear illustration of escalation. 
When one Bully increases the severity of their attack, it appears to signal to others that similar escalation is acceptable, leading additional Bullies to intensify their behavior. 
While Defender activity is present in this quadrant, particularly in $\mathcal{M}_{4,4}$ with a local prevalence of 6\% and a global prevalence of 45\%, such motifs remain limited, and the quadrant exhibits little meaningful resistance to bullying behavior. 
Overall, Quadrant IV reflects environments where bullying occurs repeatedly and with little interruption, shaping much of the global motif landscape through widespread, largely unopposed aggression.

\BfPara{Broader Behavioral Dynamics}
We now zoom out further to get a broader perspective and highlight a few significant motif structures.

\textit{Coordinated Behavior}: these manifest as \textit{star} structures, embodying two primary variants: 
(1) The \textit{Mobbing} star structure is characterized by three Bullies targeting a single Victim. 
It emerges as the most distinct and dominant structure across the dataset ($\mathcal{M}_{1}$, $\mathcal{M}_{2}$, and $\mathcal{M}_{4}$ in Fig.~\ref{fig:motifs-global}), particularly within Quadrant III ($\mathcal{M}_{3,2}$ and $\mathcal{M}_{3,3}$ in Fig.~\ref{fig:joint-plot}) and Quadrant IV ($\mathcal{M}_{4,1}$, $\mathcal{M}_{4,2}$, and $\mathcal{M}_{4,3}$). 
The primary variation of this motif, $\mathcal{M}_{1}$, is exceptionally common, exhibiting Local and Global Prevalence Scores of 11\% and 71\% respectively. 
Structurally, the Victim functions as a terminal sink node, absorbing directed aggression from three separate Bullies. This isolation suggests a contagion effect: once a Victim is marked by one Bully, they become a target for others, accelerating the harassment. 
While the most prevalent form ($\mathcal{M}_{1}$ in Fig.~\ref{fig:motifs-global}) involves low-intensity edges, the structure evolves into heavier variants ($\mathcal{M}_{2}$ and $\mathcal{M}_{4}$), indicating that as the mobbing normalizes within the group, the severity of the attacks escalates. 
(2) The \textit{Mobilization} star structure represents the inverse of the mobbing dynamic. 
Instead of multiple Bullies, there is a single Victim being pacified by multiple Defenders within the wider network. 
This structure is a defining feature of Quadrant II, particularly the motif $\mathcal{M}_{2,1}$, which appears in nearly 74\% of the quadrant’s sessions and accounts for over 10\% of the local interaction volume. 
The pattern is also visible in the global baseline ($\mathcal{M}_{6}$) and in Quadrant I ($\mathcal{M}_{1,1}$ and $\mathcal{M}_{1,4}$). 

\textit{Distributed Defense:} these are the structural inverses of the Mobbing star. 
Rather than collective aggression, these motifs represent collective resistance. 
This resistance manifests in three distinct configurations: a single Defender managing multiple threats ($\mathcal{M}_{3}$, $\mathcal{M}_{3,1}$, and $\mathcal{M}_{4,4}$), a balanced standoff ($\mathcal{M}_{9}$ and $\mathcal{M}_{3,4}$), or an overwhelming defense ($\mathcal{M}_{2,2}$ and $\mathcal{M}_{2,3}$). 
Among these, motif $\mathcal{M}_{3,1}$ is particularly dominant, appearing in 93\% of Quadrant III sessions and accounting for 15\% of the local interaction volume. Crucially, with the exception of one heavy edge in $\mathcal{M}_{2,3}$, these interactions are characterized by low-intensity edges. 
This suggests that Defenders are utilizing a deterrence-oriented strategy. 
Rather than engaging in high-conflict battles, they act as informal moderators, spreading their intervention across the group to signal behavioral boundaries and regulate aggression without causing escalation.


\textit{The Lack of Cycles:} while our analysis prioritizes significant structures, the absence of specific patterns is equally telling. 
Notably, our results show a complete lack of cycle structures. 
In network science, an acyclic graph implies a strict hierarchy or a unidirectional flow of energy from a source node to a sink node. 
In a cyberbullying context, we would typically expect to observe reciprocity, a feedback loop where a Defender pushes back against a Bully, and the Bully retaliates. 
The absence of this cycle suggests that Bullies operate mainly as hit-and-run aggressors: they attack a Victim and disengage. 
Even when a Defender intervenes, the Bully does not fight back, implying they either ignore the intervention or shift targets. 
Consequently, the network functions as a cascading waterfall where aggression flows from the Source (Bully) to the Target (Victim) and finally to the Sink (Defender), as seen in linear motifs such as $\mathcal{M}_{1,3}$. 
This acyclic nature reveals a topology of unresolved conflict; interactions are linear and fleeting, preventing the formation of the resilient, reciprocal community structures found in healthier social networks~\cite{coleman1988socialcap}.

\section{Discussion}

\subsection{Broader Impact} 
We believe our findings can help foster new proactive moderation techniques that address cyberbullying on social media platforms. 
To illustrate, our proposed Bully and Victim scores could be included in session moderation dashboards to help moderators make judgment calls on the balance of power within a given session and identify and correct problematic behavior.
For instance, if a moderator observes the Bully score in a session increasing at a faster rate than the Victim score---indicating that the session is becoming bully-dominated---they can intervene before the cyberbullying escalates further and prevent the victim from exposure to additional harm.
Potential interventions could involve temporarily banning users, locking the session to prevent new comments, and deleting comments that violate community guidelines. Moreover, because motifs, themselves, encode rich, nuanced, mesoscale user interactions that are seldom captured by conventional text-based cyberbullying detection algorithms, integrating them into training datasets can help existing models make better predictions.

\subsection{Limitations \& Future Work} 
This work is primarily exploratory in nature.
It does not utilize any supervised or unsupervised learning techniques to further extract patterns, nor does it use existing cyberbullying role classifiers~\cite{sandoval_roles} to annotate datasets and subsequently build Session Graphs from emerging social media platforms.
Furthermore, this work was conducted on an Instagram dataset containing only cyberbullying sessions.
This increases the likelihood that our findings may have resulted from confounding factors and selection bias.
In addition, this work focuses on analyzing interactions in \textit{static}, completed sessions. 
However, our proposed graph framework provides a stepping stone towards studying cyberbullying interactions as they unfold over time. 
A temporal analysis could also help identify the tendency for Bully and Victim scores to gravitate towards the origin, as seen in Fig.~\ref{fig:joint-plot}.
Finally, a notable limitation of this work stems from the lack of statistical validation of the findings.
That is, every design decision could be validated through a \textit{null model}, thus highlighting an important direction for future work. For example, one could randomly rewire the edges of a Session Graph to assess the significance of the observed Bully and Victim scores.
Future work can also explore alternative mappings of comment severity scores to the edge weights, which, by design, contribute significantly to both the Victim and Bully scores and how motifs manifest.
The systematic application of this methodology at every design decision point would increase the robustness of this approach by providing greater contextualization and facilitate a better understanding of the impact of each choice.

\section{Conclusion}
Using 414 Instagram sessions annotated for severity and cyberbullying roles, we utilize a network science framework to construct cyberbullying interaction graphs.
To quantify the balance of power between the Victims and Bullies within sessions, we proposed two metrics, the Victim Score and Bully Score. 
We found that most sessions exhibit a negative Victim Score and a slightly positive Bully Score.
Moreover, the joint distribution of the Bully and Victim Scores reveals four quadrants to explore.
These embody qualitatively distinct dynamics between Bullies and Victims---evenly-matched sessions in which Bullies and Victims engage in comparable levels of activity; sessions where the Victims and Defenders dominate; sessions in which Bullies face substantial pushback; and sessions in which Bullies dominate.
To further understand the more granular cyberbullying dynamics, we enumerated motifs---small subgraphs that provide a mesoscale view of connectivity patterns.
Through the Global (across session) and Local (within session) Prevalence scores, we identified the most frequently-occurring motifs.
The analysis of prevalent motifs highlighted emergent patterns, such as coordinated behavior (stars) and distributed defense.
To our knowledge, this is the first study that leverages human annotations at the session and comment level to explore network-level granular cyberbullying interactions on Instagram.



\begin{acks}
This work was supported by National Science Foundation Awards \#2435164, \#2435165, and \#2336386, and a Google Award for Inclusion Research.
\end{acks}

\bibliographystyle{ACM-Reference-Format}
\bibliography{references}


\end{document}